# Hardware-in-the-Loop Co-Simulation Based Validation of Power System Control Applications

Marcel Otte*, Fabian Leimgruber†, Roland Bründlinger†, Sebastian Rohjans*, Aadil Latif‡, Thomas I. Strasser†
*Hamburg University of Applied Sciences, Hamburg, Germany, email: {firstname.lastname}@haw-hamburg.de
†AIT Austrian Institute of Technology, Vienna, Austria, email: {firstname.lastname}@ait.ac.at
‡National Renewable Energy Laboratory, Golden, CO, United States, email: aadil.latif@nrel.gov

*Abstract*—Renewables are key enablers for the realization of a sustainable energy supply but grid operators and energy utilities have to mange their intermittent behavior and limited storage capabilities by ensuring the security of supply and power quality. Advanced control approaches, automation concepts, and communication technologies have the potential to address these challenges by providing new intelligent solutions and products. However, the validation of certain aspects of such smart grid systems, especially advanced control and automation concepts is still a challenge. The main aim of this work therefore is to introduce a hardware-in-the-loop co-simulation-based validation framework which allows the simulation of large-scale power networks and control solutions together with real-world components. The application of this concept to a selected voltage control example shows its applicability.

## I. Introduction

Renewable energy sources are key enablers for the realization of a sustainable energy supply [1]. However, network operators and energy utilities have to mange their intermittent behavior and limited storage capabilities by ensuring the security of supply and power quality. Advanced control approaches, automation concepts, and communication technologies have the potential to address these challenges by providing new intelligent solutions and products [2], [3]. Therefore, the traditional power system is transformed into a cyber-physical energy system, a so-called smart grid [4].

In several research projects such advanced approaches are already in development which allowing to operate the power systems with renewable generators in a more effective way compared to traditional concepts [5]–[7]. A very good example of those developments is provided by the ELECTRA IRP project which has developed an advanced control architecture, called the "Web-of-Cells" (WoC). The whole power system over all voltage levels is divided into cell-based control areas where local problems are being solved locally using the flexibility of generators and loads in those cells [6]. Different control schemes as well as corresponding observable and control functions are being developed in order to allow a much higher penetration level of renewables.

The validation of certain aspects of smart grid systems, especially advanced control and automation concepts, is still a challenge. A holistic and integrated approach for analyzing and evaluating complex configurations in a cyber-physical systems manner which supports the design, implementation, and deployment phases is still not available [8]. Advanced simulation and Hardware-in-the-Loop (HIL) approaches have the potential to overcome this situation and to provide proper validation and testing methods [9]–[11].

The main aim of this paper is to discuss a hardware-in-the-loop co-simulation-based validation framework which allows the coupling of suitable analysis and assessment tools for the evaluation of large-scale power networks and control solutions together with real-world components. This approach is currently developed in the ERIGrid project [12] and tested on a voltage control example from ELECTRA IRP [6].

Following this introduction, Section II provides an overview of the related work in validating smart grids systems with focus on simulation-based approaches. A simulation-based concept is introduced in Section III which allows the simulation of large-scale electric grids together with emulated power system and control components in a hardware-in-the-loop co-simulation manner. In order to validate the proposed approach and to show the power of it, a voltage control example from ELECTRA IRP has been implemented which is discussed in Section IV. The paper is concluded with the main findings and an outlook about the future work in Section V.

## II. Validating Smart Grid Systems and Applications

Simulations are a common tool to analyze and evaluate power system applications [13], [14]. In the past individual domains like power grids, automation and control, as well as Information and Communication Technology (ICT) have been quite often analyzed independently [8]. In order to address the needs and requirements of evaluating smart grid systems more integrated simulation approaches and concepts covering the aforementioned domains are necessary [15]. Advanced simulation approaches as a helpful tool for research and development but also for education of smart grid systems gain more attention nowadays [16]–[18]. Hybrid models addressing physics (i.e., continuous time-based) and ICT/automation-related (i.e., discrete event-based) aspects are therefore quite often in usage [10].

The coupling of domain-specific simulation models—in the literature referred to as co-simulation or co-operative simulation—is an approach for the joint simulation of models developed with different, domain-specific tools [10], [11],

[18]. During the execution of such a coupled approach intermediate results are mainly exchanged during specific points in time, where individual models are solved independently. Recent work on smart grid systems co-simulation covers power system's analysis, automation and control, ICT but also electricity market issues [11], [18].

Compared with pure software simulations, the execution of models in real-time allows also the analysis of the correct timing behavior of the system's and components under investigation. Specific simulation environments—called Digital Real-time Simulators (DRTS)—are being used in order to guarantee the correct timing behavior of the simulation models [19]. In later stages of the development process of solutions and products even more interesting is the coupling with real devices. Such a so-called HIL approach has been already used in various applications in the power and energy systems domain [9], [11], [18], [20]. Even more, the coupling of HIL concepts with co-simulations offers a lot of advantages since it allows to implement large-scale power grid models in corresponding simulators (which is not always possible in DRTS due to limited computing resources) and to couple them with real equipment (e.g., inverter-based distributed energy resources) where often detailed simulation models are quite difficult to develop. However, the coupling of co-simulation environments and DRTS together with real components is still a challenging research topic which needs further investigation and corresponding tools [8], [18].

## III. Proposed Validation Environment

As outlined above the coupling of co-simulation approaches with real components and devices has great potential but still need further research and development, especially on the coupling and tooling side. This issue is being addressed in this work where a corresponding coupling framework is introduced and discussed in the following sections.

### A. Overview of the Coupling Concept

The proposed concept of a HIL co-simulation environment which is being used for evaluating complex power system applications—as outlined in the introduction section–is mainly based on three elements. As depicted in Fig. 1, the coupling framework has to provide a real-time signal exchange between automation and control applications (referred to as "Simulation" in Fig. 1), the power grid simulation (i.e., "Grid" in Fig. 1), and hardware components coupled to it (i.e., "Hardware" in Fig. 1). It has to be noted that each component may provide different interfaces for the coupling of the tools which may also be implemented with different programming languages. Therefore, a coupling approach is required which guarantees a real-time signal exchange between the aforementioned elements. Since different simulation tools, controller hardware and software as well as real power system components (e.g., inverter-based distributed energy resources) may need to be coupled, the corresponding framework must be designed in a proper way. Synchronization and the provision of corresponding interfaces to those tools and components are the main requirements for such a framework, where a prototypical realization is introduced below.

An illustrative example where a Post Primary Voltage Control (PPVC) control scheme and corresponding functions form ELECTRA IRP is being validated with this HIL co-simulation-based approach, is provided later in this paper in Section IV.

### B. Prototypical Realization

Considering all mentioned requirements and needs for coupling, the use of a virtual bus-system as a synchronization and interfacing concept can provide a synchronized signal exchange between the coupled tools. Such a concept is realized by the so-called LabLink approach which has especially been developed for coupling and interfacing simulation tools and power system laboratory components in a flexible and extensible way. This middleware approach, with a prospective aim to be published as an open source solution, provides the possibility to connect different tools via Python or Java interfaces [21].

As outlined in Fig. 2 the core element of LabLink is a MQTT[1] Broker. LabLink clients connect to the broker and communicate with each other via the publish and subscribe paradigm. The client communication is event-based and asynchronous whereas a so-called synchronisation client (Synchronisation Host) coordinates other clients in an experiment (i.e., to have a common time base). After defining the simulation parameters in a corresponding configuration, the Synchronization Host has the specification of the required and additional clients and their requirements for the corresponding experiment.

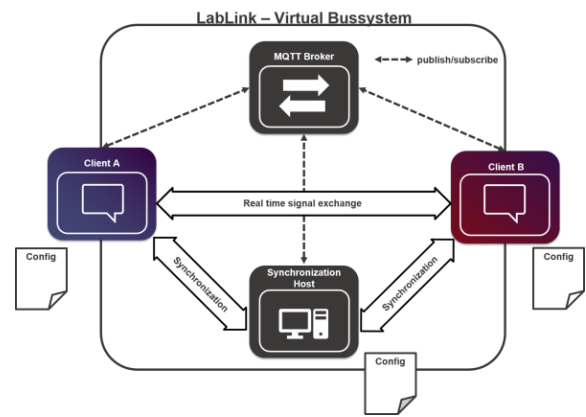

Figure 2. Realization of the coupling of tools using the LabLink framework.

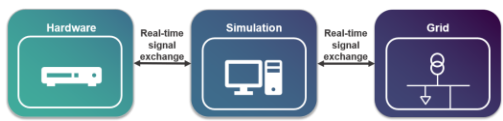

Figure 1. Concept for coupling simulators and hardware components.

[1]Message Queuing Telemetry Transport

The interface of the clients are equal, whereby only the interface for the subscriber (e.g., hardware, databases, or another software) has to be implemented. Each client is able to provide data as well as consume them from other clients. On this basis, a co-simulation validation environment has been realized in this work based on LabLink where the following five clients are being connected (see Fig. 3) in order to validate complex power system applications as outlined above:

- *Typhoon HIL Client:* Typhoon HIL as a DRTS for emulating and validating hardware components (e.g., whole converter concepts, converter control) providing software and hardware interfaces. In this work the Typhoon HIL Python API was used to exchange signals (node voltages, currents, etc.) between a large-scale power grid simulation and an emulated converter device (in Typhoon HIL DRTS) via LabLink.
- *Simulation Client:* This Client—which usually contains automation and control algorithms (e.g., cell-based voltage control in case of ELECTRA IRP)—directly uses the API of the connected power system simulator PowerFactory. Due to the fact that this API is not usable for more than two subscriber or multiple threads at the same time, the simulation and the PowerFactory Client are combined into one multithreaded LabLink client which uses the PowerFactory API in a single threaded manner.
- *PowerFactory Client:* PowerFactory supports several interfaces [20]. The Python API is an ideal fit for the LabLink environment Python bindings. This client can provide all data points (node voltages and currents, frequency, active and reactive power, etc.) of each grid component and is able to consume data from other clients (setpoints for active and reactive power to converter-based generators), which are being published via LabLink.
- *InfluxDB Client:* Stores and archives timestamped data points published by other LabLink clients (e.g., Typhoon HIL, PowerFactory). The simulation results stored in this database can be monitored, e.g., by Grafana (a web-based dashboard visualization software for timestamped metrics). Thus, a live monitoring of the power systems application can be realized.
- *Synchronization Client:* Contains all simulation parameters (duration, time-resolution, connected clients, etc.) and it differs from the other mentioned clients how it is specified and implemented. The Synchronization Client is mainly defined by configuration files containing the mentioned simulation experiment parameters.

All clients (except the Synchronization Client) have the same configuration structure which defines LabLink communication metadata.

## IV. VOLTAGE CONTROL VALIDATION EXAMPLE

With the outlined validation environment a large number of test cases are feasible. Each client of the LabLink is able to be tested. For the validation of the above outlined HIL co-simulation-based approach a selected use case from the ELECTRA IRP project is being used.

### A. Web-of-Cell Voltage Control Approach

As outlined in the introduction ELECTRA IRP targets a cell-based control approach (i.e., WoC) where various observable and control functions are being developed [6]. Within the scope of the project, the concept of the PPVC voltage control has been proposed [22]. The main idea under investigation is the possibility of defining "Cells" within a distribution network, each with its own PPVC controller that works independently. In this work cells are been defined by applying a clustering approach using the electrical distance as input.

The proposed validation framework is being used for a proof-of-concept evaluation of the voltage control approach together with a real controller device. In this experiment the CIGRE Medium Voltage (MV) distribution test grid [23] well as an emulated smart grid converter from AIT—called "ASG-Converter" [24]—together with the PPVC algorithm is been applied as outlined in Fig. 3.

### B. Identification of ELECTRA Cells

For defining the Cell Objects of the CIGRE European MV test grid, a clustering approach based on the normalized electrical distance is used [25]. For this, the calculation of the Jacobian matrix is necessary to obtain the $\frac{\delta Q}{\delta u}$ matrix by

$$\begin{bmatrix} \Delta P \\ \Delta Q \end{bmatrix} = \begin{bmatrix} \frac{\Delta P}{\Delta \delta} & \frac{\Delta P}{\Delta u} \\ \frac{\Delta Q}{\Delta \delta} & \frac{\Delta Q}{\Delta u} \end{bmatrix} \begin{bmatrix} \Delta \delta \\ \Delta u \end{bmatrix} \quad (1)$$

$$J_4 = \frac{\Delta Q}{\Delta u}. \quad (2)$$

Afterwards, by calculating the inverse of $J_4$, the sensitivity matrix B can be derived by

$$B = J_4^{-1} = \frac{\Delta u}{\Delta Q} \quad (3)$$

$$b_{ij} = \frac{\Delta u_i}{\Delta Q_{ij}}. \quad (4)$$

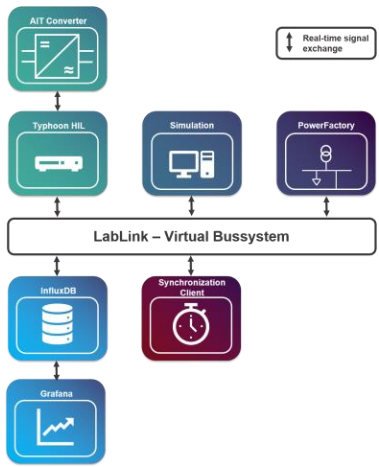

Figure 3. Simulation subscriber overview.

As a next step the attenuation matrix $a$ needs to be calculated by dividing the non-diagonal elements by the diagonal elements using the following equation

$$a_{ij} = \frac{b_{ij}}{b_{jj}}. \qquad (5)$$

For obtaining the normalized distance matrix the electrical distance matrix needs to be calculated as

$$D_{ij} = -\log(a_{ij} \cdot a_{ji}) \qquad (6)$$

$$D_{ij}^{norm} = \frac{D_{ij}}{max(D_i)}. \qquad (7)$$

By using this approach, the normalized electrical distance matrix presented in Fig. 4 shows that the CIGRE MV test feeder has strong voltage coupling. This is essentially because the line lengths at the start of the feeder are much longer than the line lengths at the end of the feeder. This means that the nodes most susceptible to over voltage have tight voltage coupling and dividing the feeder into cells might not be possible. Ensuring cells have weak voltage coupling is important as it would limit the impact of one cell's regulations on its neighbors.

One way to emphasis the utility of a cell-based approach is to modify the line lengths such that voltage zones with weak voltage interdependence are created. In this work, three line lengths have been modified (Line 1: from 2.8 km to 0.8 km, line 2: from 4.4 km to 1.4 km and line 12: from 1.3 km to 6.3 km) to create regions with weak voltage interdependence as shown in Fig. 4(b).

Results from clustering have been used to define cells within a network as depicted in Fig. 5.

*C. Test Case*

For the proof-of-concept validation experiment two parts of the environment are mainly on focus. On the one hand the power system control application (Cell-based Simulation Client) and on the other hand the emulated ASG-Converter (Typhoon HIL with ASG-Converter connection) with the real controller board. To fit the converter in to the grid, a definition of the role of the converter is necessary. Fig. 6 shows a particular feeder of the CIGRE MV network where *PV* 09 is placed in a low Voltage area of Cell One. This random chosen PV-Object will be deactivated in PowerFactory and instead the ASG-Converter emulation takes place as a HIL co-simulation. Each of these zones should have at least one controllable device.

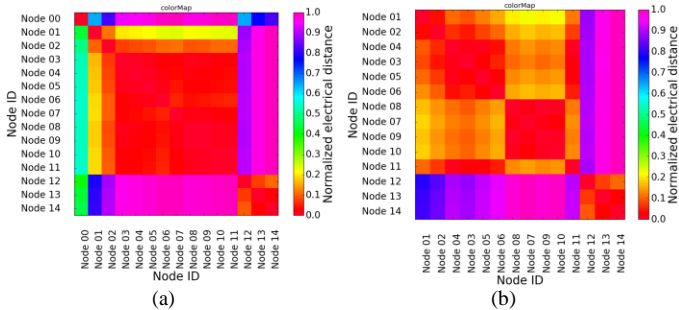

Figure 4. Color plot of the normalized electrical distance matrix for the CIGRE MV test network with (a) original and (b) modified line length.

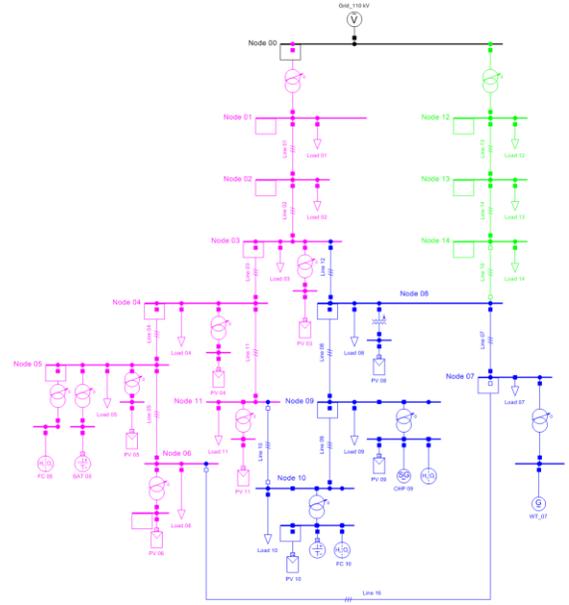

Figure 5. PowerFactory representation of the CIGRE MV test network divided into three cells (i.e., Cell 1 in cyan, Cell 2 in blue, Cell 3 in green).

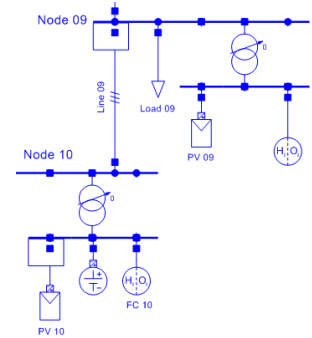

Figure 6. Extract of the Cell-based grid with *P V* 09.

The DC side (Photovoltaic) of the converter needs to be replaced by a Python script. The corresponding real-time simulation model in Typhoon HIL, which is connect via LabLink and the corresponding clients is depicted in Fig. 7. The model contains the PV array, DC measurements, the DC-AC converter, the interface device, grid measurements, and parts of the test feeder which is connected to the grid simulation in PowerFactory. Moreover, the real converter controller (i.e., embedded controller) is connected to the DRTS.

One of the advantages of this set-up is the possibility to connect the real controller board to the simulation instead of deriving a less precise component model for the grid simulation which may need a lot of resources to be developed.

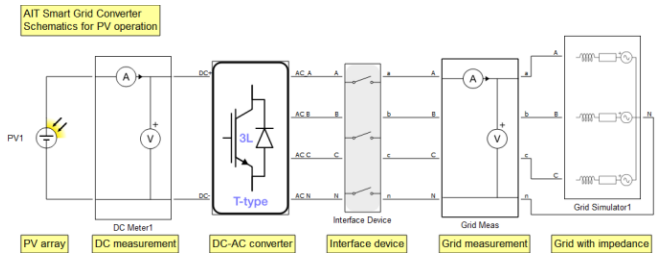

Figure 7. Converter model in Typhoon HIL.

### D. Realized Test

As described in Section III-B the Synchronization Client needs all the simulation parameter for the Synchronization Host. The implemented test includes a one day steady-state simulation within one minute step-size. All mentioned clients from Fig. 3 are required and no additional clients are implemented. Furthermore, the ASG-Converter will have a signal exchange via LabLink with Node 21 of the grid; i.e., voltage set-points from PowerFactory simulation are communicated to Tyhoon HIL where the grid connection is modeled as voltage source and the later provides active and reactive power back. Due to the balanced configuration of the three phases, only one voltage data-point of the grid is necessary. A "Cell" object is created for each zone defined in the PowerFactory project file. Each Cell "Object" is created for each controllable device connected within the zone. Properties for each object are populated (required for optimization). The cell controller implemented in Python use external solver (Differential Evolution available in SciPy library). PowerFactory has been mainly used as a load flow engine for this proof-of-concept validation.

### E. Achieved Results and Discussion

During the execution of the experiment the behavior of the mentioned cell-based power system control application was monitored. The test confirms the real-time signal exchange between the clients. Regarding to the SCADA-system in Fig. 8 the overview of the Typhoon HIL proves that the voltage of the PowerFactory simulation grid successfully achieved the

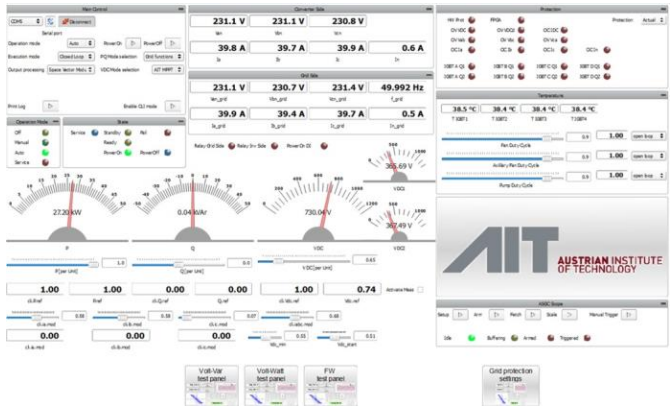

Figure 8. SCADA UI for controlling the ASG-Converter.

Typhoon HIL (data-point "Van_Grid" in section "Grid Side") After sending the signals to the ASG-Converter, the Typhoon HIL DRTS received the signals from the converter (i.e., the active and reactive power are the "P" and "Q" values on the illustrated power display).

As a further result the database (i.e., InfluxDB Client) consumes all data-points, which are published via LabLink. With the second monitoring software (i.e., Grafana), which is directly connected to the database (see Fig. 3), the whole simulation signal exchange is displayed. In Fig. 9, the voltage (first graph) and the reactive and active power (second graph in load-reference arrow system) are displayed. The voltage of the node reaches up to 238 Volt. Thereby the $Q = f(U)$ control algorithm of the ASG-Converter provides a phase shift. As a further proof, the Synchronization Host guarantees the real-time signal exchange, which can be seen in the live monitoring as well. On this occasion, long time periods (approximately 30 s) after 15 simulation steps are also monitored, which are based on the global numerical optimization of the cell-based simulation and has, due to the Synchronization Host, no effect on the real-time signal exchange.

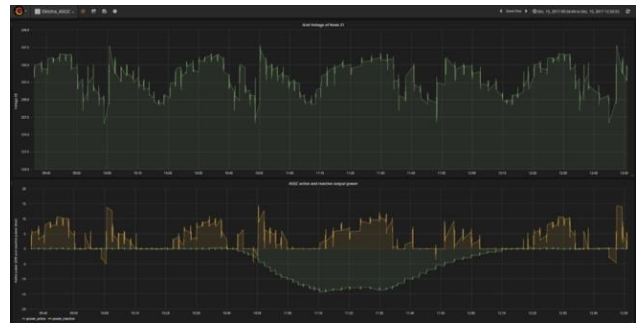

Figure 9. Live monitoring of values from the experiment.

Furthermore, the WoC voltage control approach leads to the achievement of a comparison between the different amount of ELECTRA cells. As presented in Fig. 10, the network losses can be reduced by defining more cells, including the PPVC controller. Also the impact of the line length shows a big influence on the normalized electrical distance approach for these cells. As outlined in Fig. 10(b) the normalized losses were able to improved by a maximum change of 3.30 %.

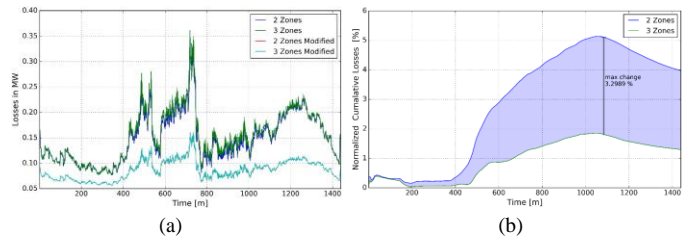

Figure 10. Network losses for the three scenarios: (a) total losses, (b) Normalized losses and unmodified

*F. Lessons Learned*

As a summarized test result, the validation demonstrates a high number of possibilities and powerful flexibility for the simulation environment. Several LabLink Clients can participate in the co-simulation, whereby the complexity of a large-scale communication model will be simplified. As mentioned before the cell-based power system application delivers several achievements. Which means the clustering approach using the normalized electrical distance provides a suitable tool for the identification of ELECTRA cells. Furthermore, the coupled simulation and HIL experiment shows the successful operation of the ELECTRA voltage control in context of the WoC approach. Decrease in loses can be observed for the used testing scenario compared to the base case and the ELECTRA voltage control approach is also suitable to be used for traditional distribution grid optimization. For this validation case, real-time co-simulation with HIL coupling was a very helpful tool for carrying out the proof-of-concept evaluation using a real component controller.

## V. Conclusions

In this work, a validation framework has been presented, which allows the proof-of-concept evaluation of future power system control applications. Within this environment several co-simulations are feasible and the flexibility of the LabLink coupling approach has shown, that new subscribers are easy to realize. Based on an interesting control concept—ELECTRA WoC—a test case was successfully implemented.

Summarizing, the proposed validation framework reduces the effort for prospective simulations with hardware and software components. New requirements are easier to realize and the real-time settings are given by the LabLink.

For the future additional, more complex simulation examples with larger test grids and additional clients are planned which are addressing scalability issues. Moreover, also the protoypical realization of the HIL co-simulation-based validation environment needs further improvement.

## Acknowledgment

This work is partly supported by the European Community's Seventh Framework Program (FP7/2007-2013) under project "ELECTRA IRP" (Grant Agreement No. 609687) as well as by the European Community's Horizon 2020 Program (H2020/2014-2020) under project "ERIGrid" (Grant Agreement No. 654113). Further information is available at the corresponding websites electrairp.eu and erigrid.eu.

## References

[1] M. Liserre, T. Sauter, and J. Hung, "Future Energy Systems: Integrating Renewable Energy Sources into the Smart Power Grid Through Industrial Electronics," *IEEE Industrial Electronics Magazine*, vol. 4, no. 1, pp. 18–37, 2010.

[2] V. Gungor, D. Sahin, T. Kocak, S. Ergut *et al.*, "Smart Grid Technologies: Communication Technologies and Standards," *IEEE Transactions on Industrial Informatics*, vol. 7, no. 4, pp. 529–539, 2011.

[3] T. Strasser, F. Andren, J. Kathan, C. Cecati *et al.*, "A review of architectures and concepts for intelligence in future electric energy systems," *IEEE Transactions on Industrial Electronics*, vol. 62, no. 4, pp. 2424–2438, 2015.

[4] H. Farhangi, "The path of the smart grid," *IEEE Power and Energy Magazine*, vol. 8, no. 1, pp. 18–28, 2010.

[5] G. Kariniotakis, L. Martini, C. Caerts, H. Brunner, and N. Retiere, "Challenges, innovative architectures and control strategies for future networks: the web-of-cells, fractal grids and other concepts," *CIRED - Open Access Proceedings Journal*, vol. 2017, pp. 2149–2152, 2017.

[6] L. Martini, H. Brunner, E. Rodriguez, C. Caerts *et al.*, "Grid of the future and the need for a decentralised control architecture: the web-of-cells concept," *CIRED – Open Access Proceedings Journal*, vol. 2017, no. 1, pp. 1162–1166, 2017.

[7] N. Retiere, G. Muratore, G. Kariniotakis, A. Michiorri *et al.*, "Fractal grid – towards the future smart grid," *CIRED – Open Access Proceedings Journal*, vol. 2017, pp. 2296–2299, 2017.

[8] T. Strasser, F. Pröstl Andren, G. Lauss, R. Bründlinger *et al.*, "Towards holistic power distribution system validation and testing-an overview and discussion of different possibilities," *e & i Elektrotechnik und Informationstechnik*, vol. 134, no. 1, pp. 71–77, 2017.

[9] X. Guillaud, M. O. Faruque, A. Teninge, A. H. Hariri *et al.*, "Applications of real-time simulation technologies in power and energy systems," *IEEE Power and Energy Technology Systems Journal*, vol. 2, no. 3, pp. 103–115, 2015.

[10] P. Palensky, A. van der Meer, C. Lopez, A. Joseph, and K. Pan, "Cosimulation of intelligent power systems: Fundamentals, software architecture, numerics, and coupling," *IEEE Industrial Electronics Magazine*, vol. 11, no. 1, pp. 34–50, 2017.

[11] ——, "Applied cosimulation of intelligent power systems: Implementing hybrid simulators for complex power systems," *IEEE Industrial Electronics Magazine*, vol. 11, no. 2, pp. 6–21, 2017.

[12] T. Strasser, C. Moyo, R. Bründlinger, S. Lehnhoff *et al.*, "An integrated research infrastructure for validating cyber-physical energy systems," in *Industrial Applications of Holonic and Multi-Agent Systems*. Springer International Publishing, 2017, pp. 157–170.

[13] F. M. Gonzalez-Longatt and J. L. Rueda, *PowerFactory applications for power system analysis*. Springer, 2014.

[14] J. A. Momoh, *Electric power system applications of optimization*. CRC press, 2017.

[15] S. Rohjans, S. Lehnhoff, S. Schütte, F. Andrén, and T. Strasser, "Requirements for smart grid simulation tools," in *2014 IEEE 23rd International Symposium on Industrial Electronics (ISIE)*, 2014, pp. 1730–1736.

[16] M. Kezunovic, "Teaching the smart grid fundamentals using modeling, simulation, and hands-on laboratory experiments," in *2010 IEEE Power and Energy Society General Meeting*, 2010, pp. 1–6.

[17] R. Podmore and M. R. Robinson, "The role of simulators for smart grid development," *IEEE Transactions on Smart Grid*, vol. 1, no. 2, pp. 205–212, 2010.

[18] C. Steinbrink, S. Lehnhoff, S. Rohjans, T. I. Strasser *et al.*, "Simulation-based validation of smart grids–status quo and future research trends," in *International Conference on Industrial Applications of Holonic and Multi-Agent Systems*. Springer, 2017, pp. 171–185.

[19] M. O. Faruque, T. Strasser, G. Lauss *et al.*, "Real-time simulation technologies for power systems design, testing, and analysis," *IEEE Power Energy Techn. Syst. Journal*, vol. 2, no. 2, pp. 63–73, 2015.

[20] M. Stifter, F. Andrén, R. Schwalbe, and W. Tremmel, "Interfacing power-factory: Co-simulation, real-time simulation and controller hardware-in-the-loop applications," in *PowerFactory Applications for Power System Analysis*. Germany: Springer Verlag, 2015, pp. 343–366.

[21] D. Stahleder, D. Reihs, M. Nöhrer, and F. Lehfuss, "Lablink – a novel co-simulation tool for the evaluation of large scale ev penetration focusing on local energy communities," *to be presented in CIRED Workshop 2018*, June 7-8, Ljubljana, Slovenia, 2018.

[22] J. Merino, J. Rodríguez-Seco, I. García-Villalba, A. Temiz *et al.*, "Electra irp voltage control strategy for enhancing power system stability in future grid architectures," *CIRED – Open Access Proceedings Journal*, vol. 2017, pp. 1068–1072, 2017.

[23] Task Force C6.04, "Benchmark systems for network integration of renewable and distributed energy resources," CIGRE, Tech. Rep. 575, 2014.

[24] R. Bründlinger, R. Ablinger, and Z. Miletic, "Ait smart grid converter(sgc) controller featuring sunspec protocol support utilizing hardware-in-the-loop (hil) technology," SunSpec Meeting, Sep. 2016.

[25] A. Latif, I. Ahmad, P. Palensky, and W. Gawlik, "Zone based optimal reactive power dispatch in smart distribution network using distributed generation," in *2017 Workshop on Modeling and Simulation of Cyber-Physical Energy Systems (MSCPES)*, 2017.